\documentclass[twocolumn,showpacs,showkeys,pre]{revtex4}
\usepackage{amssymb}

\usepackage{graphicx}
\usepackage{hyperref}

\begin{document}

\title{The size distribution, scaling properties and spatial organization of urban clusters: a global and regional perspective}
\author{\surname{Till} Fluschnik, \surname{Steffen} Kriewald, \surname{Anselmo} Garc\'ia Cant\'u Ros, 
\surname{Bin} Zhou, \surname{Dominik E.} Reusser, \surname{J\"urgen} P. Kropp, \surname{Diego} Rybski}
\affiliation{Potsdam Institute for Climate Impact Research (PIK)}
\email{anselmo@pik-potsdam.de}
\begin{abstract}
Human development has far-reaching impacts on the surface of the globe. 
The transformation of natural land cover occurs in different forms and 
urban growth is one of the most eminent transformative processes. 
We analyze global land cover data and extract cities as defined by 
maximally connected urban clusters. 
The analysis of the city size distribution for all cities 
on the globe confirms Zipf's law.
Moreover, by investigating the percolation properties of the clustering of 
urban areas we assess the closeness to criticality for various countries.
At the critical thresholds, the urban land cover of the countries 
undergoes a transition from separated clusters to a gigantic component on 
the country scale.
We study the Zipf-exponents as a function of the closeness to percolation 
and find a systematic decrease with increasing scale, 
which could be the reason for deviating exponents reported in literature. 
Moreover, we investigate the average size of the clusters as a function 
of the proximity to percolation and find country specific behavior.
By relating the standard deviation and the average of cluster sizes -- 
analogous to Taylor's law -- we suggest an alternative way to identify 
the percolation transition.
We calculate spatial correlations of the urban land cover and 
find long-range correlations.
Finally, by relating the areas of cities with population figures we address the 
global aspect of the allometry of cities, finding an exponent $\delta\approx 0.85$, 
i.e.\ large cities have lower densities.
\end{abstract}

\maketitle

\section{Introduction}
\label{sec:Introduction}
In the beginning of the last century, F.~Auerbach \citep{AuerbachF1913}
claimed "The law of population concentration". 
In various phases \citep{RybskiD2013}, the seemingly scale-invariant character 
of city size distributions is most often described in terms of a power-law
\begin{equation}
p(X)\sim X^{-\zeta} 
\enspace ,
\label{eq:zipf}
\end{equation}
where $p$ denotes the probability density of observing within a scoped region a
city sample of size $X$. For this expression empirical estimations of the
exponent $\zeta$ closely deviate around 2 -- 
so called \emph{Zipf's law for cities}, 
after G.K.~Zipf's \citep{ZipfGK1949}. 

While several city growth models have been proven successful in reconstructing 
power-law city size distributions
\citep{GibratR1931,SimonHA1955,MakseABHS1998,RybskiRK2013},
statistical tests have also assigned a great plausibility to alternative
functional forms, as in the case of log-normal distributions \citep{EeckhoutJ2004}.

In this work we estimate the global city size distribution, based on both, 
urban land cover and population. 
For this purpose we apply an orthodromic 
version of the recently proposed \emph{City Clustering Algorithm} (CCA)
\citep{RozenfeldRABSM2008}, to account for a more accurate estimation of 
the areas of all urban settlements of the world (approximately $250,000$). 
We find that Zipf's law approximately holds to a great extent for city areas 
and to a lesser extent for urban population.

As a matter of fact, characterization of the spatial organization and scaling
properties of urban clusters depends on the definition of a city boundary. In
particular, defining a city boundary by means of the CCA requires to specifying
a distance below which adjacent urban areas are considered to be part of the 
same cluster. 
The variation of this parameter involves a problem similar to percolation
transition; beyond a critical clustering distance value, a giant cluster 
component emerges. 
We explore further the influence of the choice of the
clustering parameter on the spatial organization and scaling properties of
urban land cover clusters, for several European countries. 

This paper is organized as follows: In Sec.~\ref{sec:ccalc} we provide a brief
description of the CCA algorithm and of the land cover and population databases.
Section~\ref{sec:GCsize} reports on the global city size distribution for city
area and population. In Sec.~\ref{sec:SizeandPerc} we present the country scale
results; Sec.~\ref{subsec:perctrans} addresses the CCA percolation 
transition; Sec.~\ref{ssec:size} elaborates on the connection between
the scaling properties of city size distributions and the CCA percolation 
transition; in Sec.~\ref{ssec:avemass} we discuss the scaling of the average
size of city clusters, approaching the percolation transition; in
Sec.~\ref{ssec:Taylor} we show that the variability of city cluster sizes
also exhibits scaling, in the form of the so called Taylor's law; regarding the
spatial organization of city clusters, in Sec.~\ref{ssec:spacor} we present results
on the scaling of spatial correlations. Finally, in Sec.~\ref{sec:Allom} we
explore the global aspect of the city allometry relationship, i.e., the
power-law relation between population and area for approximately 70,000 
cities under scope. 
The main results of this work are summarized and discussed in
Sec.~\ref{sec:Summ}.

\section{City clustering and land cover data}
\label{sec:ccalc}
Since a city might include natural gaps, such as the River Thames in London or
other topographic obstacles, it is convenient to define cities as 
connected clusters of neighboring populated sites. 
This idea has been recently implemented in the so called 
\emph{City Clustering Algorithm} (CCA) \citep{RozenfeldRABSM2008}, 
which is an adapted version of the more general Burning Algorithm 
\citep{StaufferA1994}.

Basically, CCA identifies any pair of adjacent urban spatial units (either by
population or land cover) as belonging to the same urban cluster if these are 
located within a distance $l$ from each other. Thus, when applied to an entire
region, CCA provides a mean to determine the areas and boundaries of the cities
contained within, according to the parameter $l$, which represents 
a degree of coarse-graining. 

At the global scale, data on the spatial distribution of population is only
available for administrative boundaries or as raster data with a rather coarse
resolution. Therefore, we opted for determining the area of cities a remote
sensing based classification of land cover data, as provided by the
\emph{GlobCover 2009 land cover map} \citep{GlobCover2009} at a grid resolution
of approx.~$0.308$\,km (at the Equator). From the 23 land cover classes, we
selected and aggregated those corresponding to urban land use.

\begin{figure*}[t]
\centering
\includegraphics[width=\textwidth]{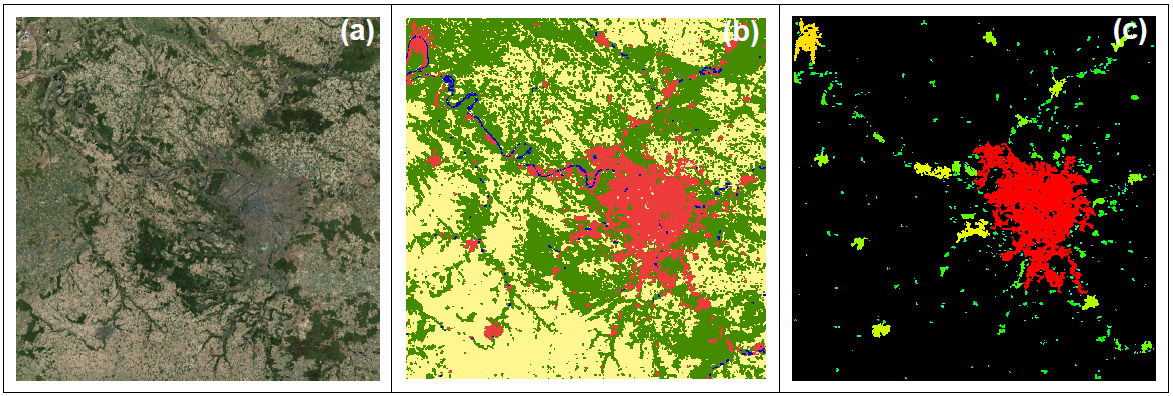}
\caption{Application of city clustering to urban land cover data. 
The following panels illustrate different aspects of the city of Paris and its surroundings:
(a) Remote sensing image as extracted from the ArcGIS~10 component ArcMap.
(b) Urban land cover data as obtained from the GlobCover 2009 land cover map. 
The colors indicate urban (red, class 190), water bodies (blue, class 210), 
forests and grasslands (green, classes 20-110), 
and rainfield croplands (yellow, class 14).
(c) From the urban land cover and by taking $l=4$\,km, the identified clusters 
are color coded according to the logarithm of their size: 
from small (light blue) via medium (green) to large (red).
The cutout has the approximate area of ($215$\,km)$^2$.} 
\label{fig:illustrative}
\end{figure*}

For the sake of illustration, Fig.~\ref{fig:illustrative} exhibits the
application of the CCA to the land cover data for Paris and its surroundings. 
A satellite image of the region is displayed in Fig.~\ref{fig:illustrative}(a),
the corresponding aggregated land cover classes in
Fig.~\ref{fig:illustrative}(b), and the urban clusters identified after
application of the CCA in panel Fig.~\ref{fig:illustrative}(c).
 
Since the raster cell size decreases from the Equator to the poles, the use of
the Euclidian metric is not suitable for the application of CCA at the global
scale. Accordingly, orthodromic distances were considered in a new
implementation 
of the CCA in order to provide a more accurate representation of distances 
and areas across different latitudes on a 
sphere. In the orthodromic 
representation, a distance determined in terms of 
the latitude~$y_i$ and longitude~$x_i$ coordinates is given by 
$d_{i,j} = R_{{\rm Earth}} \cos^{-1}(\psi_{i,j})$ where
$\psi_{i,j}=\left( \sin(x_i)\sin(x_j) + \cos(x_i)\cos(x_j)\cos(y_i-y_j) \right)$
with the radius of the terrestrial sphere~$R_{{\rm Earth}}\approx 6.371 
\times 10^3$\,km.

\section{Global city size distribution}
\label{sec:GCsize}

\begin{figure}[t]
\centering
\includegraphics[width=\columnwidth]{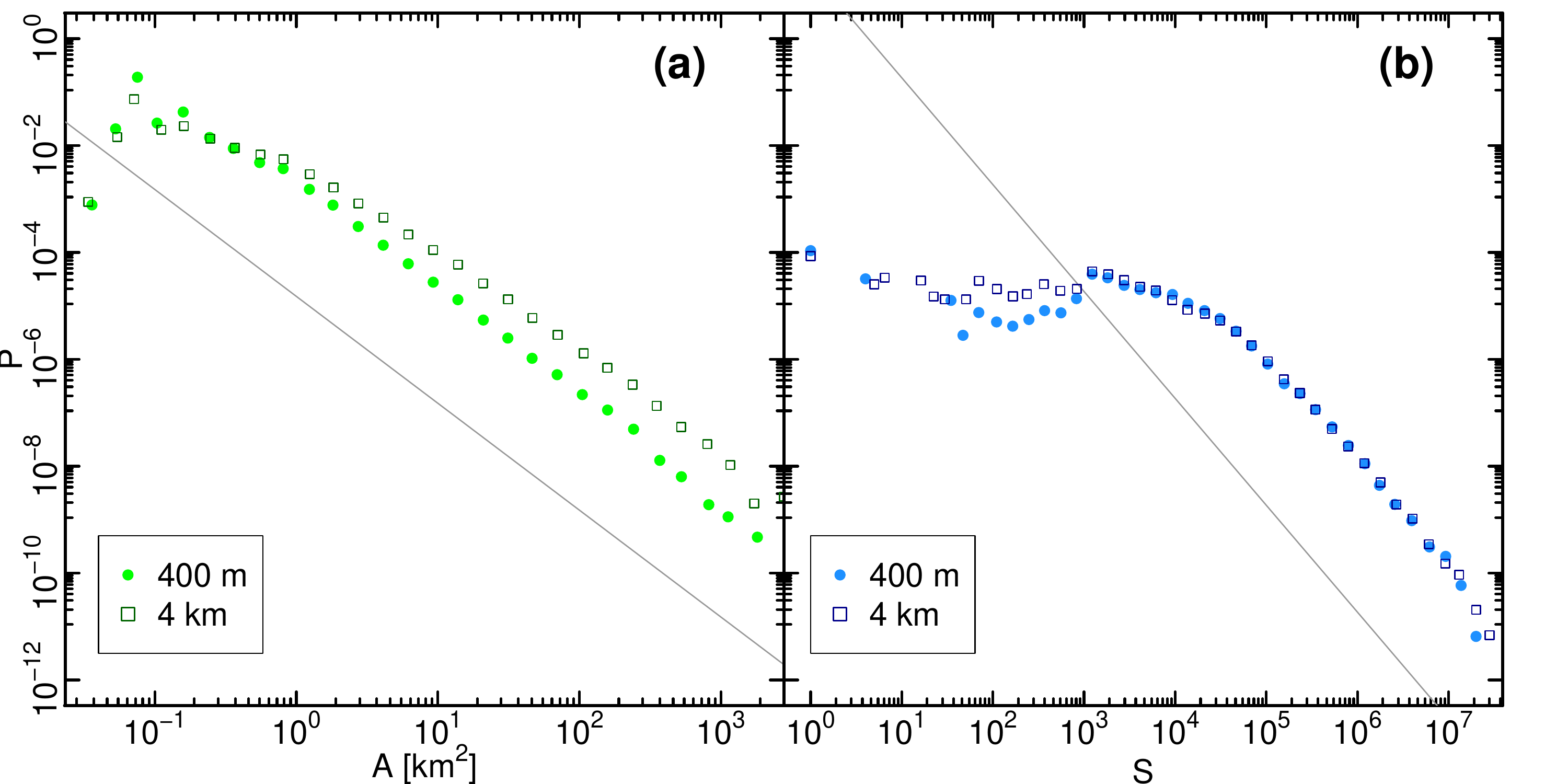}
\caption{
Probability density of city size in terms of area and population.
(a) Cluster area distribution $p(A)$, as obtained by applying the 
City Clustering Algorithm (CCA) to global land cover data and 
extracting all urban clusters on the globe.
For $A>0.1$\,km$^2$ we estimate 
$\zeta_A \approx 1.93$ for $l=0.4$\,km (249,512 clusters) and 
$\zeta_A \approx 1.75$ for $l=4$\,km (46,754 clusters).
(b) Cluster population distribution $p(S)$, as obtained from 
associating population settlement points with the	 
urban clusters identified by means of CCA.
For $S>10^4$ we estimate 
$\zeta_S\approx 1.85$ for $l=0.4$\,km and 
$\zeta_S\approx 1.75$ for $l=4$\,km. 
In both panels: $l=0.4$\,km (circles), $l=4$\,km (squares).
The solid grey lines have slope $-2$.
}
\label{fig:zipfglobal}
\end{figure}

With the aim of addressing the global city size distribution, we applied
CCA with a clustering distance of $l=0.4km$ to the entire global land cover
database referred to in Sec.~\ref{sec:ccalc}, from which we extracted $249,512$
urban clusters. The resulting area probability density $p(A)$ is shown in
Fig.~\ref{fig:zipfglobal}(a). Besides deviations for small sizes -- which
are mainly due to the discreteness of the grid cells -- we find a fair power-law
relation in agreement with Eq.~(\ref{eq:zipf}), 
with $X=A$ and $\zeta_A\approx 1.93$ for $A\ge 1$\,km$^2$. 

In terms of population, the global probability density, 
was obtained using population data from the 
\emph{Global Rural Urban Mapping Project} (GRUMP) \citep{GSP2011},
which comprises coordinates, names, and population figures of $67,935$
administrative units (estimated data for the year 2000). 
From this sample we found $16,908$ urban settlement points which are located 
inside an urban cluster or within a distance of $l=0.4$\,km 
(following a similar approach as in \citep{RozenfeldRGM2011}).
Figure~\ref{fig:zipfglobal}(b) shows the population probability
density $p(S)$. Since, the number of small clusters which could be
assigned to a cluster and accordingly to a number of inhabitants 
is small, we observe in Fig.~\ref{fig:zipfglobal}(b) deviation 
from the power law distribution Eq.~(\ref{eq:zipf}) at the lower end. 
Therefore, the power law fitting is carried out for urban
clusters above $10^4$ inhabitants, 
resulting in an exponent $\zeta_S=1.85$. 
Accordingly, we observe that Zipf's law approximately holds for 
the cities on the global scale, 
whereas the actual exponent is smaller than $2$.
We explore also $l=4$\,km and similar power-law size distributions, 
however with a different exponent in the case of the areas 
(Fig.~\ref{fig:zipfglobal}(a)).

Studies of global population city size distributions were reported in
\citep{ZanetteM1997,BattyM2008}, for a reduced subset of cities, e.g.\ the
$2,700$ largest clusters in the case of the former. 
Distributions of city size in terms of \emph{area} has
been considered previously, e.g.\ in
\citep{SchweitzerS1998,MakseABHS1998,RozenfeldRGM2011,KinoshitaKIY2008,ArcauteHFYJB2013} 
at the regional and country scale. 
More recently, a global analysis has positively tested Zipf's law by
considering temporally stable night lights as a proxy indicator for 
human habitation and anthropogenic land use \citep{SmallC}.

\section{Percolation transition and size distribution on the country scale}
\label{sec:SizeandPerc}

\subsection{Percolation transition}
\label{subsec:perctrans}

It is worth noting that when the clustering parameter $l$ is set to a very small
value the CCA does not take any effect, in the sense that the urban clusters
thereby identified correspond trivially to those observed from the input land
cover map. On the other hand, in the opposite limit of very large $l$, most of
the urban area under scope are assigned to a same giant cluster component. 
Accordingly, when applied to a large area, for intermediate values of $l$ it is
expectable to observe a percolation transition of the urban clusters. 
As it turns out, it becomes natural to inquire into this possibility
and to eventually address the spatial properties drawn from application of the
CCA in the light of concepts and methods stemming from percolation theory
\citep{StaufferA1994,BundeH1991}.

At the country scale, let us address the possible percolation transition that
may occur at the level of the urban patch clustering when changing the 
parameter $l$ in a typical application of the CCA -- which, more in general,
constitutes a problem inherent to the ambiguous character of the definition of 
city boundaries \citep{RozenfeldRABSM2008,BerryOK2012,ArcauteHFYJB2013}. It is
in order to mention that the scale defined by the clustering parameter $l$
determines the type of percolation transition under scope -- for small $l$,
the transition resembles the one occurring in site percolation on a square
lattice, while for large $l$ it can be further assimilated to the one observed
in continuum percolation problems \citep{BundeH1991}. Here, we are interested in
the value $l_{\rm c}$ at which the giant cluster component spans within a
country territory. The critical value $l_{\rm c}$ is analogous to the critical
occupation probability $P_{\rm c}$, which constitutes the control parameter in
most of lattice percolation formulations \citep{BundeH1991}. 
Both quantities are approximately related by 
$P\sim l^\beta$ with $\beta\approx 2$. 

In real world data, however, it can be difficult to identify such an $l_{\rm c}$
percolation threshold. We find that the average cluster size excluding the
largest cluster, $\langle A\rangle^*$, constitutes a sensitive indicator of the
transition. In infinite systems, $\langle A\rangle^*$ diverges at $P_{\rm c}$
\citep{BundeH1991} -- in case of finite systems a (finite) peak occurs.
Similarly, one can in principle detect the presence of a peak in $\langle
A\rangle^*$ around a value $l_{\rm c}$ when applying CCA to the urban land cover
of different countries. Since in the limit of small $l$ the urban clusters
identified by the CCA approximate the cells of urban land cover, 
we conjecture that a small $l_{\rm c}$ value constitutes a proxy indicator of
the percolation threshold of the urban land cover.

For illustrative purposes, let us consider the case of Austria. 
Figure~\ref{fig:perccol}(a) depicts the plot of $\langle A\rangle^*$ vs.\ $l$. 
As it can be observed, for $l<l_{\rm c}$ the average cluster size increases 
strongly with~$l$, yet gradually, and it drops sharply for $l>l_{\rm c}$. In
the case of Austria we find that the peak occurs at $l_{\rm c}=15$\,km. 

In a model based on correlated percolation \citep{MakseHS1995,MakseABHS1998} 
the urban/non-urban structure is formed from spatial correlations, 
i.e. the probabilities of two sites being urban/non-urban are
more similar the closer they are. 
Furthermore, a radial decay of density around the city center is assumed.
A similar approach has been recently applied to reproduce the scaling 
properties observed in urban land parcels \citep{BitnerHF2009}.
The dynamics and characteristics of the percolation transition of
the urban land cover has been investigated also by means of diffusion
limited \citep{Murcio} and gravity based \citep{RybskiRK2013} stochastic
aggregation models of city growth.

\subsection{City size distribution}
\label{ssec:size}

Let us now consider the influence of the coarse-graining used in defining a city
cluster, i.e. parameter $l$ in the CCA, on the scaling of the city size
distributions. 
At this stage, we stress the fact that for many countries $l_{\rm c}$ cannot be
identified unambiguously (see e.g.\ inset of Fig.\ref{fig:taylor}(b)), as for
instance in the presence of multiple peaks -- often a signature of large
clusters being disconnected by vastly extended topographic heterogeneities. 
Therefore, we focus on a selected set of countries exhibiting (i) a clear
percolation threshold and (ii) a large number of urban areas.

For a given $l$ value we extract all city cluster areas $A_i$ and estimate the
corresponding exponent $\zeta_A$ by applying the method proposed in
\citep{ClausetSN2009} and testing power-law against log-normal. Thus, we
first quantify the pointwise log-likelihood ratios between the fitted power-law
and fitted log-normal distributions and then apply the Voung test for non-nested
models \citep{Vuong1989}. This test essentially consists in testing the
hypothesis that both distributions are equally far away from the true
distribution, against the two alternative cases where either of each
distributions is closer to the true distribution than the other one. 
Consequently, we account only for those cases where the fitted $\zeta_A$-values 
results in positive Voung test and where the associated one-sided p-value
exceeded $0.9$ (which corresponds to a significance level of
$10$\%). For this procedure we used the corresponding R code 
(available at \url{http://tuvalu.santafe.edu/\textasciitilde aaronc/powerlaws/}). We vary $l$
and repeat the procedure to obtain $\zeta(l)$.

\begin{figure}[t]
\centering
\includegraphics[width=0.75\columnwidth]{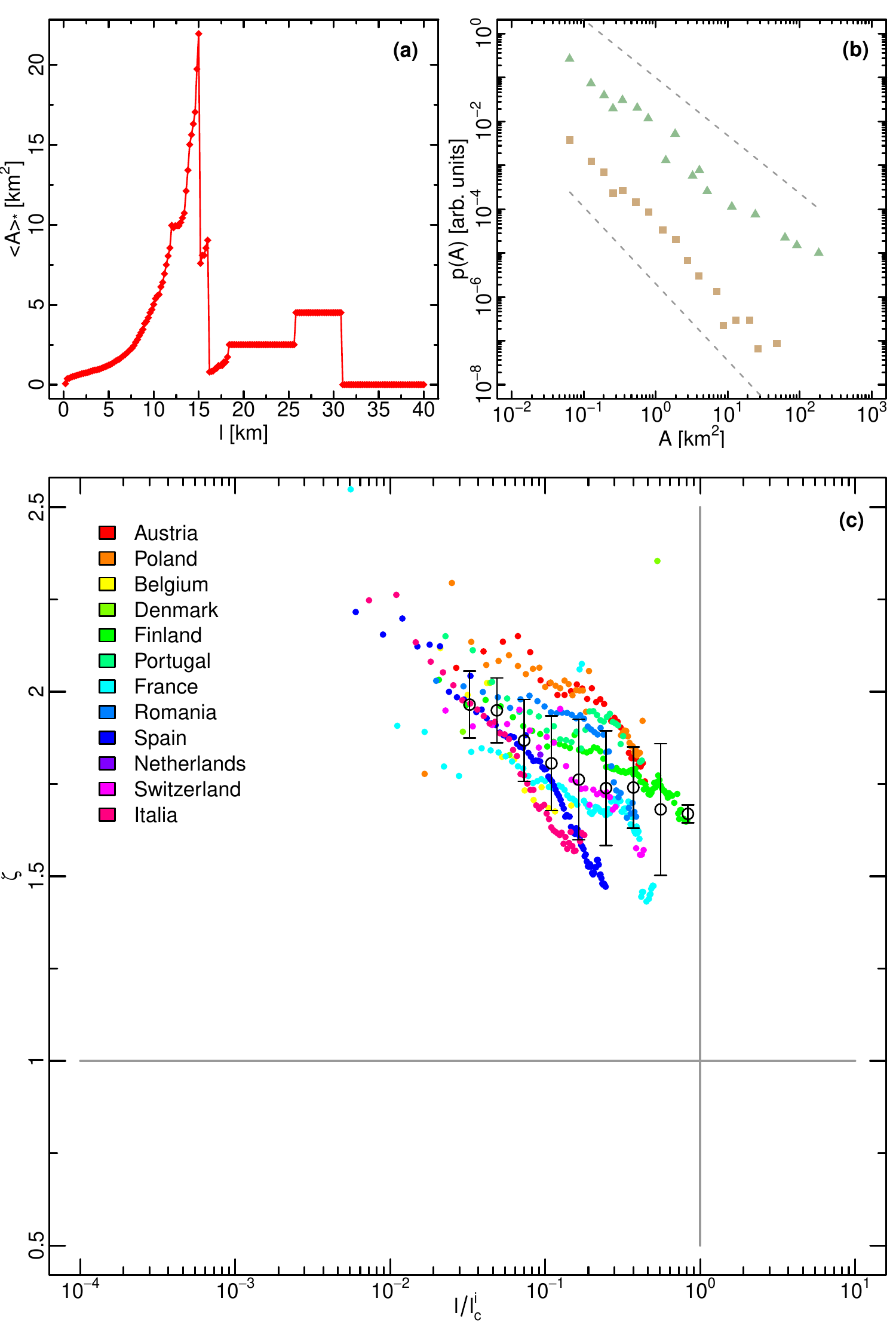}
\caption{
Percolation and Zipf's law. 
(a) Average cluster size excluding the largest component 
$\langle A\rangle^*$ as a function of the clustering parameter $l$ 
for Austria. 
The maximum is located at the percolation transition, 
which in this example is $l_{\rm c}\simeq 15$\,km. 
(b) Probability density of cluster areas $p(A)$ for Austria and for
$l\simeq\frac{1}{3}l_{\rm c}$ (370 clusters, green triangles) as well as 
$l\simeq\frac{2}{3}l_{\rm c}$ (87 clusters, brown squares).
The dotted grey lines have the slopes $-1.71$ and $-1.27$.
(c) Estimated power-law distribution exponent $\zeta_A$ as a function of 
the rescaled clustering parameter $l/l_{\rm c}$ for various countries 
as indicated by colored dots.
Since we found out that the method proposed in \citep{ClausetSN2009} has a significant deviation 
from the real value for input with less than 100 entries, we estimated the power-law distribution exponents for each country just for those $l$ with at 
least 100 clusters remaining.
The open circles represent averages in logarithmic bins and their 
error bars the corresponding standard deviations.
The exponent decreases with increasing $l/l_{\rm c}$ and 
takes values between 
$\zeta_A\approx 2$ for $l/l_{\rm c}\ll 1$ and 
$\zeta_A\approx 3/2 \dots 1$ for $l/l_{\rm c}\rightarrow 1$.}
\label{fig:perccol}
\end{figure}

Figure~\ref{fig:perccol}(b) shows the probability density $p(A)$, as obtained
for the same country as in Fig.~\ref{fig:perccol}(a) for two different values of
$l$ (for illustrative purposes, normalized histograms with logarithmic binning 
are shown). 
The fitting results in $\zeta_A=1.71$ and $\zeta_A=1.27$, 
for $l=5$\,km and $l=10$\,km, 
respectively. 
Similar decreases in $\zeta$ are also found for other countries 
(see Fig.~\ref{fig:perccol}(c)). 
From these fndings, we conjecture an approximately logarithmic dependence on the
ratio $l/l_c$, with $\zeta_A\approx 2$ for $l\ll l_c$ and 
$\zeta_A\approx 3/2 \dots 1$ for $l\rightarrow l_c$. 
We cannot determine the exact value for $l\rightarrow l_c$, 
since the sample sizes become small and the estimated $\zeta$ unreliable. 
Note that the curves in Fig.~\ref{fig:perccol}(c) do not fully collapse, 
in the sense of \citep{StanleyHE1999}, 
i.e.\ the curves do not fall on the identical line, 
from which we learn that there must be other influences beyond our analysis, 
such as heterogeneities in the urban land cover.
Another possible explanation for this could be measurement errors 
in the estimation of $\zeta_A$ and $l_{\rm c}$.
We stress the fact that the size distributions of urban land cover clusters
appear to agree with the functional form in Eq.~(\ref{eq:zipf}), independently
of the distance to the percolation threshold $l_{\rm c}$ -- in contrast to the
case of uncorrelated percolation where a power-law size distribution of
clusters emerges only in the close vicinity of the transition threshold
\citep{BundeH1991}.
 
Decreasing $\zeta$ with increasing $l$ has also been reported for the USA
\citep{RozenfeldRGM2011}.
Moreover, a recent study of a ``gravity'' based urban growth
model \citep{RybskiRK2013} has shown that 
$\zeta(P)\approx a+b\ln(P)+c\ln(1-P)$, 
where $P$ represents the site occupation probability. 
Depending on the values of $a$, $b$, and $c$, 
this expression leads to a similar decay as in Fig.~\ref{fig:perccol}(c). 
This similarity suggests a generic influence of the proximity
to the percolation threshold on the power-law size distribution 
of urban land cover.

\subsection{Average size scaling}
\label{ssec:avemass}

According to percolation theory, the average cluster size of finite clusters, 
$\langle A\rangle^*$, i.e.\ disregarding the largest cluster, 
scales with the proximity of the occupation probability to the critical 
probability, 
$\langle A\rangle^* \sim |P-P_{\rm c}|^{-\gamma}$, 
where the exponent $\gamma$ is universal and only depends on the dimension 
\citep{BundeH1991}. 
It is of our interest to explore whether the percolation of the urban land cover
clustering exhibits a similar scaling. 
Since in our analysis only few clusters
remain above the percolation transition, we omit the case $l>l_{\rm c}$ and
study $\langle A\rangle^*$ as a function of $(l_{\rm c}-l)$.

\begin{figure}[t]
\centering
\includegraphics[width=\columnwidth]{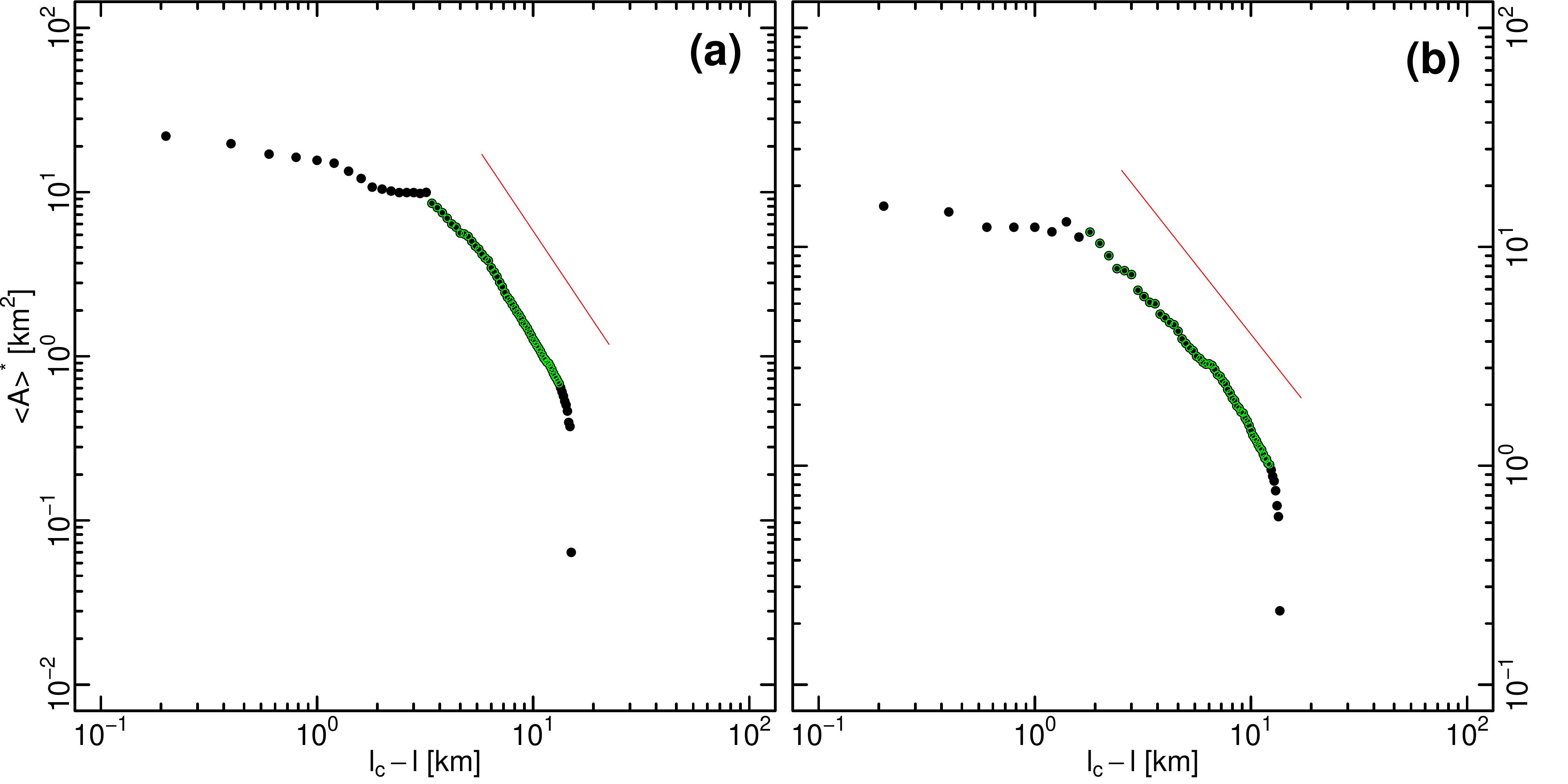}
\caption{
Fitting of the average size scaling for Austria (a) and Denmark (b). 
In both cases we have a unique clear peak in the curve of $\langle A\rangle^*$ 
against $l$. 
The red lines represent the fittings of the function $f(x)=c\cdot x^a$ on the
green highlighted parts. 
Fittings yields the parameter $a\approx-1.9674$ for Austria and 
$a=-1.2519$ for Denmark.
}
\label{fig:amscaling}
\end{figure}

Figure~\ref{fig:amscaling} shows the results for Austria and Denmark. 
Due to the finite size of the countries, $\langle A\rangle^*$ does not 
diverge for $(l_{\rm c}-l)\rightarrow 0$ and we see a plateau. 
In the other limit, $\langle A\rangle^* \rightarrow 1$ 
for large $l\rightarrow 1$.
In between we find a regime approximately following a power-law 
\begin{equation}
\langle A\rangle^* \sim (l_{\rm c}-l)^{-\gamma}
\label{eq:amscaling}
\enspace . 
\end{equation}
In general, we did not find a universal behavior as in random percolation, in
the sense that the values obtained for $\gamma$ can strongly differ among
countries. For instance, for Austria, least squares fitting provides
$\gamma\approx 2$ and for Denmark $\gamma\approx 1.25$. Such a variability in
the $\gamma$ values can result from measurement errors in the identification of
$l_{\rm c}$ (as discussed in Sec.~\ref{subsec:perctrans}) or due to systematic
structural influences occurring at larger scale, such as the presence of spatial
correlations or an accidented topography. Moreover, in some countries the plot
of $\langle A\rangle^*$ vs.\ $l$ does not exhibit a clear power law relation. On
one hand, the log-log plot of $\langle A\rangle^*$ vs $l$ can appear as composed
by many linear segments, or, furthermore, the presence of a power-law-like
segment cannot even be adequately prescribed over a substantial range of
$l$-values.

\subsection{Taylor's law for city size distribution}
\label{ssec:Taylor}

Beyond the behavior of the average size with $l$, characterizing the scaling
properties of urban clusters requires also to attend to statistical 
regularities occurring at the level of the variability of the cluster sizes. For
this purpose we elaborate on an empirical relation first established in the
context of ecology, the so called Taylor's law \citep{TaylorLR1961,SmithHF1938}.
In systems satisfying Taylor's law, the standard deviation and the average of a
quantity are related by a power-law. Both quantities are either temporal or over
ensembles. According to \cite{MenezesB2004}, in the case of temporal variability
it follows either a linear or square-root scaling. For a recent review on this
topic we refer to \citep{EislerBK2008}.

In the case of cities, we consider the standard deviations and average of
cluster sizes for a given $l$, i.e.\ $\sigma^*_A(l)$ and $\langle
A\rangle^*(l)$, whereas we omit the largest cluster 
(this is necessary since at least for $l>l_{\rm c}$ it is an outlier). 
By varying $l$, the ensemble of cluster sizes and the hypothesized power-law can
be investigated.
\begin{equation}
\sigma^*_A\sim(\langle A\rangle^*)^\alpha
\label{eq:taylor}
\end{equation}

\begin{figure}[t]
\centering
\includegraphics[width=\columnwidth]{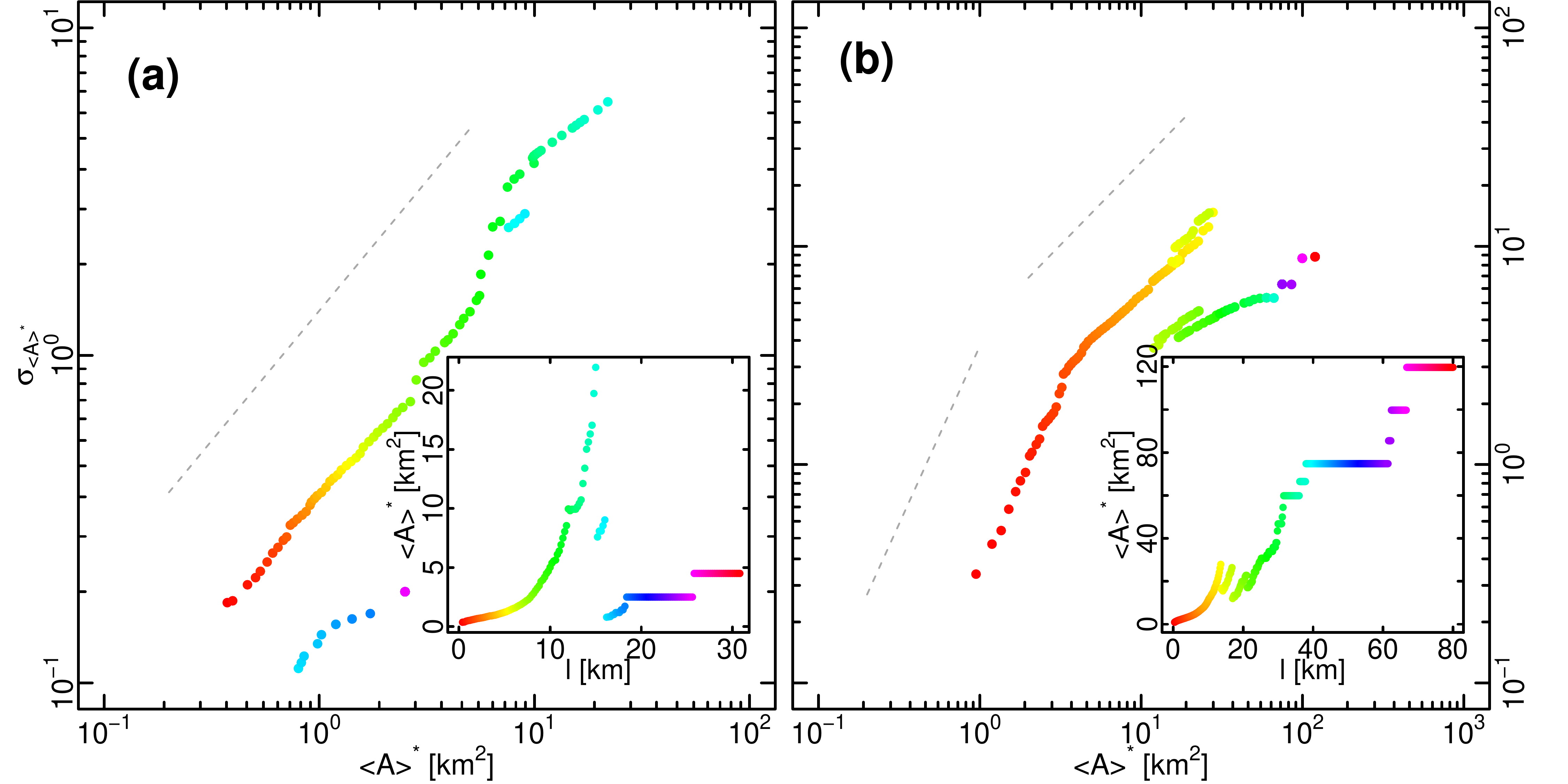}
\caption{
Taylor's law for city sizes. 
The standard deviation $\sigma_A^*$ of cluster sizes disregarding 
the largest cluster is plotted as a function of $\langle A\rangle^*$ 
for various values of $l$. 
The panels show the results for (a) Austria and (b) Spain.
The insets depict the corresponding curve of $\langle A\rangle^*$ against $l$
(see also Fig.~\ref{fig:perccol}(a)). 
In the panels and insets corresponding coloring is used 
in order to enable comparison. 
For Austria, the standard deviation $\sigma_A^*$ 
reaches its maximum at $l=15\,km$ (light-blue) and 
for Spain it is located at $l=13.6\,km$ (yellow).
In the former case this maximum corresponds to the percolation point. 
In the latter case the maximum corresponds to the first peak one can 
identify in the curve of $\langle A\rangle^*$ against $l$.
}
\label{fig:taylor}
\end{figure}

In Fig.~\ref{fig:taylor} we show the results for 
two example countries (Austria and Spain). 
While the major panels display $\sigma^*_A$ vs.\ $\langle A\rangle^*$, 
the corresponding $l$ can be inferred from the color-coded insets.
For Austria (Fig.~\ref{fig:taylor}(a)), a power-law regime is found 
for small $l$, i.e.\ $l\leq 10$\,km with $\alpha \approx 0.79$. 
The slope seems to hold up to $l \approx 15$\,km 
but separated by jumps in the standard deviation.
In contrast, for Spain (Fig.~\ref{fig:taylor}(b)), two different 
power-law regimes can be seen, 
the first up to $l \approx 4$\,km with $\alpha_1 \approx 1.63$ and 
the second one up to $l \approx 16$\,km with $\alpha_2 \approx 0.76$.
For other countries we obtained similar results. 

We conclude that Taylor's law holds to some extent 
for city sizes but there is no unique exponent and the scaling 
regimes are country specific.
Nevertheless, we observe a characteristic maximum in the 
plot of $\sigma^*_A$ vs.\ $\langle A\rangle^*$. 
In the case of Austria, this maximum matches with the percolation threshold
$l_{\rm c}$. 
In the case of Spain, the maximum standard deviation is located at
the similar position as a small peak in the representation of 
$\langle A\rangle^*$ vs.\ $l$ (inset of Fig.~\ref{fig:taylor}(b)). 
This suggests a relation between the percolation threshold $l_{\rm c}$ and
Taylor's law, where the presence of a maximum of the latter could constitute 
a mean to identify the former.

\subsection{Spatial correlations}
\label{ssec:spacor}

In the context of the analysis of the scaling properties of city clusters, it is
worth stressing the role dynamic processes underlying city growth play.
As shown in \citep{MakseHS1995,MakseABHS1998}, city cluster size distributions
are influenced by the presence of spatial correlations. The above mentioned
gravity based model of urban growth \citep{RybskiRK2013} has illustrated the
relation between the degree of compactness of urban clusters and the
exponent~$\zeta$ of the cluster size distribution Eq.~(\ref{eq:zipf}). 

With the aim of addressing the spatial organization of real urban clusters we
calculate the auto-covariance function
\begin{equation}
C(d)=
\langle (x_i-\langle x\rangle) (x_j-\langle x\rangle) | d\rangle_{i,j}
\enspace ,
\label{eq:corfunc}
\end{equation}

Here, $x_i$, $x_j$ represent the land cover of sites $i$, $j$, respectively, 
i.e.\ $x=1$ for urban and $x=0$ otherwise. 
The indices $i$, $j$ run over all land cells and the average (denoted by
brackets) is taken on those cells lying within a distance $d$, which is
predefined by logarithmic bins.

\begin{figure}[t]
\centering
\includegraphics[width=\columnwidth]{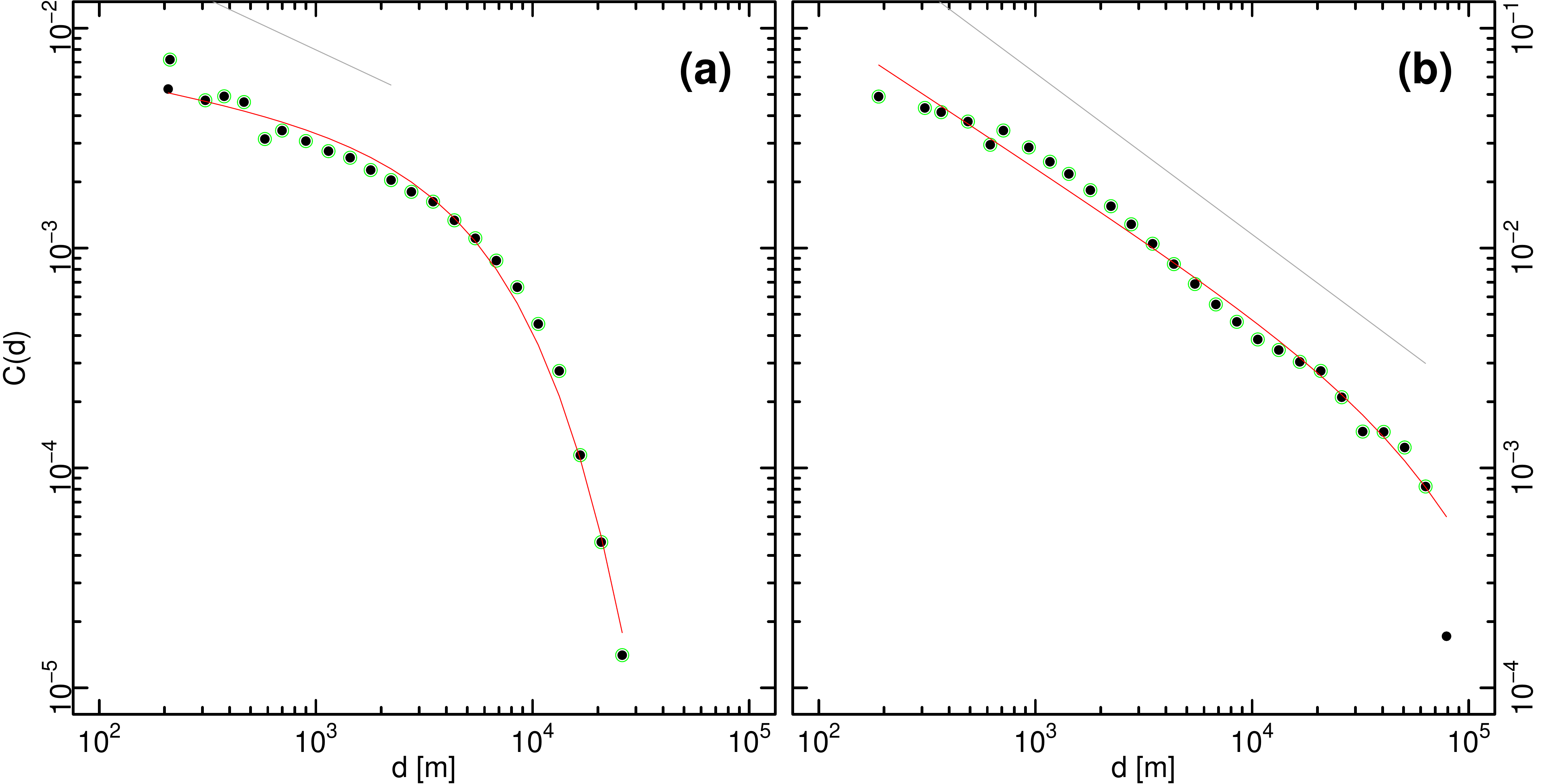}
\caption{
Spatial correlation computations for Austria (a) and Netherlands (b). The fitted
curve (red) on the green-highlighted points which follows the function $f(x):=
c\cdot x^a\cdot \exp(b\cdot x)$ (power-law with exponential cut-off) has the
values for Austria $(a,b,c)\approx(-0.177,-1.866*10^{-4},0.013)$
and for the Netherlands $(a,b,c)\approx(-0.647,-1.052\times 10^{-5},2.021)$. The grey lines
(fitted on the corresponding points below the displayed part) has slopes
$\approx -0.4615$ for Austria and $\approx -0.7331$ for the Netherlands. 
Above the shown range, $C(d)$ fluctuates around zero.
}
\label{fig:spatcor}
\end{figure}

In Fig.~\ref{fig:spatcor} we show $C(d)$ for Austria and the Netherlands as 
illustrative cases. As can be observed, $C(d)$ remains positive for scales
at least up $100$\,km and it decays with distance, approximately following a 
power-law 
\begin{equation}
C(d)\sim d^{-\xi}
\enspace .
\label{eq:corfuncpl}
\end{equation}
Note however that at large distances, the decay exhibits a cut-off where
$C(d)$ drops considerably. 
In order to take the cut-off into account, we elaborate further on the fit
\begin{equation}
C(d)\sim {\rm e}^{-\lambda d} d^{-\xi'}
\label{eq:corfplco}
\end{equation}
as used e.g.\ by \cite{ClausetSN2009} in different contexts. 

We fit Eq.~(\ref{eq:corfuncpl}) to the approximately linear regime of $\ln C$
vs.\ $\ln d$ by means of least squares and Eq.~(\ref{eq:corfplco}) by employing
non-linear curve-fitting applying the Gauss-Newton-Algorithm 
(cf. \citep{fitting}). 
While both approaches, Eq.~(\ref{eq:corfuncpl}) and Eq.~(\ref{eq:corfplco}),
lead to different exponents~$\xi$ and~$\xi'$, these are both lower than $2$, 
thereby indicating the presence of long-range correlations. 

Regarding the relation between the correlation decay and the percolation
threshold, it has been shown that long-range correlations can influence the
percolation properties \citep{WeinribA1984}. However, according with
\citep{PrakashHSS1992}, the influence of correlations on the threshold value is
only minor. For instance, for site percolation on a square lattice
\citep{PrakashHSS1992}, the site occupation probability threshold has been
shown to vary slightly between 
$P_{\rm c}\simeq 0.593$ for the case of no long-range correlation effect 
($\xi=2$ in Eq.~(\ref{eq:corfuncpl})) and 
$P_{\rm c}\rightarrow\, \approx 0.5$ for the case of strong long-range 
correlations ($\xi=0$ in Eq.~(\ref{eq:corfuncpl})). 
As mentioned in Sec.~\ref{ssec:size}, spatial inhomogeneities can hinder
unambiguous identification of the clustering threshold $l_{\rm c}$. Since
long-range spatial correlations can entail spatial inhomogeneities,
quantification the weak influence of correlations on the value of $l_{\rm c}$
cannot be achieved in many cases.

\section{City allometry -- relating area and population}
\label{sec:Allom}

An important aspect in the analysis of cities is their allometry properties, 
i.e.\ the relation between the city sizes (e.g.\ given by population) and their
socio-economic ``functions'' \citep{BettencourtW2010} and structure. Of
particular interest in this context is the relation between city size and area
\citep{BattyM2011}. Thus, we last address the relation between
urban cluster population and area. For this purpose, we combine urban land
cover data, as provided by GlobCover, with the pointwise population 
numbers of the $67,935$ administrative cities included in the GRUMP database. 
We match population and area of clusters by
summing up for each cluster those population numbers, that
are located within the distance $l$ to the considered cluster (similar to the
procedure followed in \citep{RozenfeldRGM2011}). 

\begin{figure}[t]
\centering
\includegraphics[width=\columnwidth]{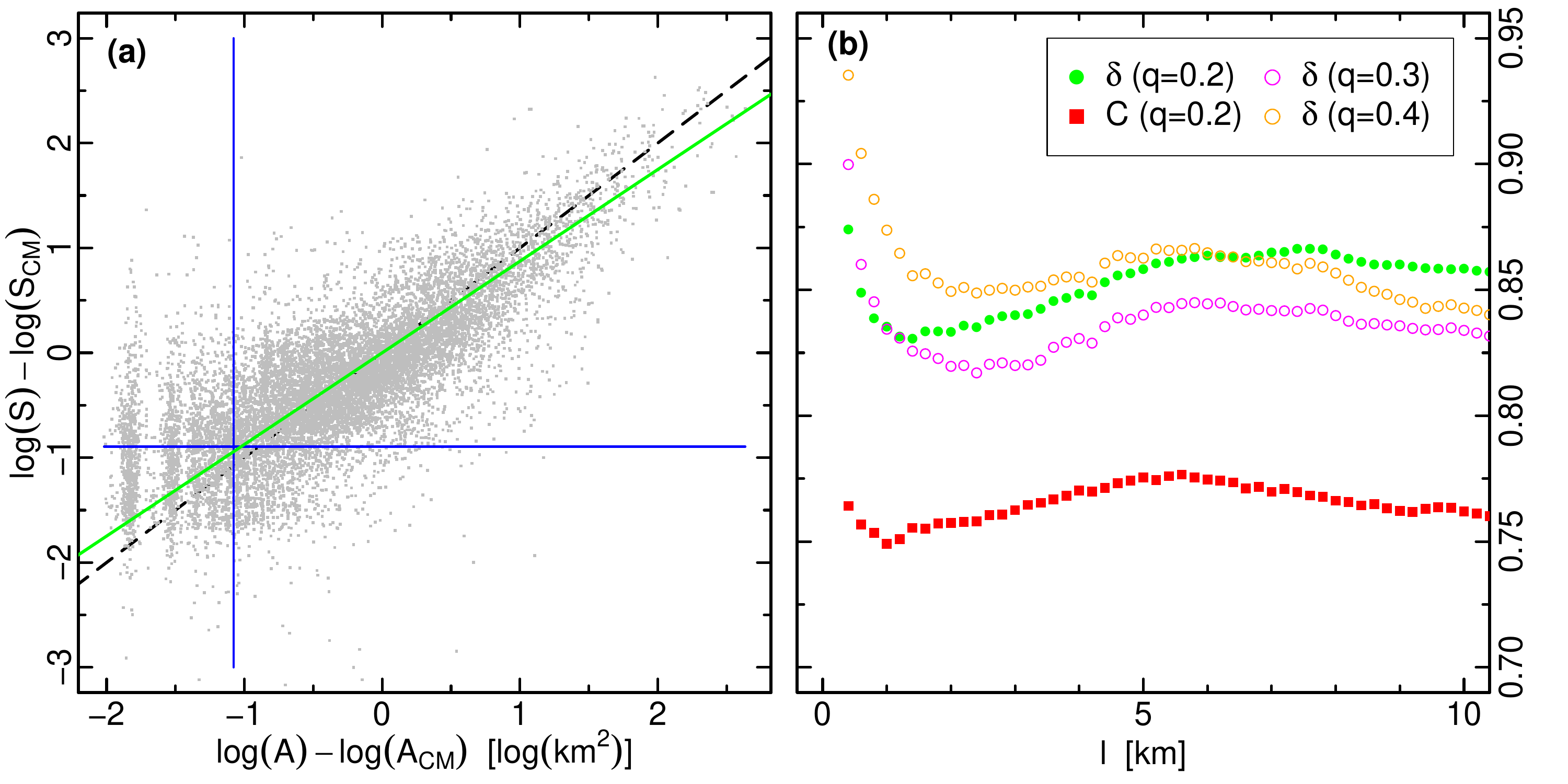}
\caption{Correlations between area and population.
(a) $\log(S)$ vs.\ $\log(A)$ for $l=0.4$\,km and of all land cover clusters 
that could be matched with population figures (grey dots).
The blue vertical and horizontal lines truncate the fraction $q=0.2$ along 
both axis in order to avoid the discreteness at small $A$. 
The green solid line corresponds to the main axis around which the momentum of inertia of 
the truncated cloud is minimal. 
In this case, $8786$ out of $12321$ clusters remain. 
It's slope $\delta$ is smaller than the diagonal 
(black dashed line, background).
(b) Slope $\delta$ (circles) and Pearson correlation coefficients $C$ (squares)
vs.\ clustering parameter $l$ for the cut-off 
$q=0.2$ (green, red), 
$q=0.3$ (magenta),
$q=0.4$ (orange).
The exponent $\delta$ is found roughly between $0.82$ and $0.87$ except 
for small $l$ where $\delta$ up to $0.87$ and $0.93$ are achieved but always 
below $1$.
For $q=0.2$ the correlations exhibit a maximum around $5$\,km.
}
\label{fig:SvsA}
\end{figure}

Figure~\ref{fig:SvsA}(a) depicts the obtained relation between area and
population. Most clusters are spread around $S\sim A^\delta$, except clusters
with small population which exhibit discreteness (stemming from the land cover
resolution) and deviate from the power-law. In order to overcome this
difficulty, we remove a fraction $q$ of clusters for both, $S$ and $A$ low
values, as indicated by blue lines in Fig.~\ref{fig:SvsA}(a). It is also
necessary to consider that a direct fitting of $\log(A)$ vs.\ $\log(S)$ and
$\log(S)$ vs.\ $\log(A)$ statistically leads to different results. 
Therefore, we assimilate $\delta$ to the slope of the longitudinal
principal axis of rotation of the cloud of points in the $\log(A)$ vs.
$\log(S)$ plot, as obtained from the eigenvector analysis of the corresponding
tensor of ``inertia''. 

In order to address the dependency of the correlations between area and
population on the observational scale, i.e.\ on the clustering parameter, in
Fig.~\ref{fig:SvsA}(b) we plot for $q=0.2$ the Pearson correlation values as a
function of $l$ and consistently find values above $0.85$ with a maximum of
approx.\ $0.88$ at approx.\ $5$\,km. Figure~\ref{fig:SvsA}(b) also shows the
values of $\delta$. For very small $l$, highest values are found between
$\delta\approx 0.85$ for $q=0.2$ and $\delta\approx 0.93$ for $q=0.4$. For
other values of $l$, the slope fluctuates but generally $\delta$ is roughly 
within $0.82$ and $0.87$. As it turns out, the results show a sublinear
population to area relation, i.e. $\delta<1$. In other words, we find that for a
given increase in population, the associated increase in area is greater for 
large cities than for small ones. 

While \citep{BattyM2011} suggests an evolution of $\delta$, 
i.e.\ that estimations of $\delta$ are decreasing since the 1940s, 
our results indicate an $l$-dependence of $\delta$, i.e.\ that the value depends
on the observational scale, but generally in the lower range of those listed in
\citep{BattyM2011}. This is a relevant fact, since, as before mentioned,
allometry could be responsible for the scaling of socio-economic quantities
with the population of cities, including urban CO$_2$-emissions
\citep{RybskiSRFK2013,NewmanK1990}.

\section{Summary and discussion}
\label{sec:Summ}

In this paper we have elaborated on the influence exerted by the degree of
coarsening resolution that is inherent to the definition of city boundaries, on
a set of indicators of the scaling, spatial organization, and allometry aspects
of urban clusters. For this purpose we implemented a version of the City
Clustering Algorithm that takes the curvature of the globe into account and
apply it to global satellite based information on the global urban land cover,
in combination with pointwise information of populated units worldwide. 

Our results show that, at the global scale, Zipf's law is found to
approximately hold to a great extent for the areas of urban clusters 
and to a lesser extent for the corresponding population. 
A shortcoming is the error introduced by automatized identification of urban
areas from satellite imagines as inherent in the land cover data used in this
study. As a matter of fact, the climatological, vegetational and structural
variety of urban clusters on the globe can hinder their classification, e.g.\
in GlobCover data Kabul city is not classified as urban. 
For a recent alternative analysis of size distributions of city clusters on the
global scale we refer to \citep{MakseGlobal}. 

At the country scale, we addressed the percolation transition that may
occur, at the level of the urban cluster definition, when coarsening the 
resolution considered to identify urban clusters. This information is relevant
as a proxy indicator for detecting the proximity to the percolation transition
occurring eventually on the real land cover of large urbanized areas -- which in
turn could exert influence on important factors of the sustainability, such as
Urban Heat Island effect (see e.g.\ \citep{ZhouRK2013}), landscape fragmentation,
water surface run-off and floods control, among others. 
As a matter, identifying an urban clustering percolation threshold can become a
cumbersome task, due, for instance, to spatial inhomogeneities in the
distribution of urban clusters at the large country scale. 

In this work we used as an indicator the behavior of the
average cluster size, excluding the largest one, which itself provides a
suitable characterization of urban clusters, for different levels of
coarsening-resolution. This analysis was applied to selected country cases, for
which we addressed the linkage between the scaling of urban cluster sizes and
the degree of resolution used for their definition. Our results appear to be at
a certain degree consistent with recent results obtained numerically in a simple
gravity-based urban growth model. Beyond the scaling and average size,
characterization of urban clusters requires addressing the scaling of deviations
of cluster sizes around the average. In particular, for the selected country
cases, we illustrate the validity of Taylor's law between average urban cluster
sizes and their standard deviations. Moreover, we show that strong deviations
from Taylor's law can be readily used to identify threshold values in proxy
indicators of the percolation transition of urban clustering, in cases where the
average cluster size fails. 

Regarding the spatial organization of urban
clusters in selected country cases, we find the presence of long-range
correlations decaying as a power law with exponential cut-off. However, for
densely urbanized areas we find that a power-law decay is further
significative, as compared to the fitting of a power-law with exponential cut
off. 

Finally, our results on the relation of area to population indicate that for a
given increase in population, the associated increase in area is greater for
large cities than for small ones. A shortcoming in our analysis of city
allometry is the minor time period discrepancy between the population and land
cover databases -- that actually represent the most updated and accurate
publicly available global databases.

\section*{Acknowledgments}
D. Rybski acknowledges E.~Arcaute and A.P.~Paolo Masucci for useful
discussions. 
This work has been funded by the Federal Ministry of Education and Research
(BMBF) through the program "Spitzenforschung und Innovation in den Neuen
L\"{a}nden" (contract "Potsdam Research Cluster for Georisk Analysis,
Environmental Change and Sustainability" D.1.1) and partly by the RAMSES
project of the European Commission under the 7th Framework Programme.

\newpage

\end{document}